\documentclass[
    aps,
    prb,
    twocolumn,
    superscriptaddress,
]{revtex4-2}

\usepackage{amsmath,amssymb,bm}
\usepackage{booktabs}
\usepackage[acronym]{glossaries}
\usepackage{graphicx}
\usepackage{listings}
\usepackage{siunitx}
\usepackage{pythonhighlight}
\usepackage[colorlinks]{hyperref}

\lstdefinestyle{Python}{
    language     = Python,
    basicstyle   = \footnotesize\ttfamily,
    commentstyle = \color{olive}\ttfamily,
    emph        = [1]{DataContainer, SAXSProjectorCUDA, SphericalHarmonics,GradientResidualCalculator,SquaredLoss,L2Norm,LBFGS},
    emphstyle   = [1]{\color{blue}\bfseries},
    emph        = [2]{use_gpu,ell_max,name,regularizer,regularization_weight,maxiter},
    emphstyle   = [2]{\color{purple}},
    emph        = [3]{add_regularizer,optimize,run_optical_flow_alignment,shifts_to_geometry,run_mitra},
    emphstyle   = [3]{\color{teal}\bfseries},
    numberstyle = \color{orange},
    stringstyle = \color{red},
    frame       = lines,
    rulecolor   = \color{gray},
}

\hypersetup{
    linkcolor=black,
    citecolor=black,
    filecolor=black,
    urlcolor=black,
}

\setacronymstyle{long-short}
\newacronym{ct}{CT}{computed tomography}
\newacronym{dd}{DD}{discrete directions}
\newacronym{ddf}{DDF}{directional dark-field}
\newacronym{dti}{DTI}{diffusion tensor imaging}
\newacronym{fa}{FA}{fractional anisotropy}
\newacronym{fbp}{FBP}{filtered backprojection}
\newacronym{gk}{GK}{Gaussian kernels}
\newacronym{mitra}{MITRA}{modular iterative tomographic reconstruction algorithm}
\newacronym{mri}{MRI}{magnetic resonance imaging}
\newacronym{motr}{MOTR}{momentum total variation reconstruction}
\newacronym{nn}{NN}{nearest neighbors}
\newacronym{radtt}{RADTT}{robust and denoised tensor tomography}
\newacronym{rsm}{RSM}{reciprocal space map}
\newacronym{saxs}{SAXS}{small-angle x-ray scattering}
\newacronym{sh}{SH}{spherical harmonics}
\newacronym{sigtt}{SIGTT}{spherical integral geometric tensor tomography}
\newacronym{sirt}{SIRT}{simultaneous iterative reconstruction technique}
\newacronym{tt}{TT}{tensor tomography}
\newacronym{waxs}{WAXS}{wide-angle x-ray scattering}
\newacronym{xrd}{XRD}{x-ray diffraction}
\newacronym{zh}{ZH}{zonal harmonics}

\DeclareMathOperator*{\argmin}{argmin}

\renewcommand{\vec}[1]{\boldsymbol{#1}}
\renewcommand{\matrix}[1]{\mathbf{#1}}
\newcommand{\mumott}{\textsc{Mumott}}
\newcommand{\numba}{\textsc{numba}}

\newcommand{\chalmersphys}{%
    Department of Physics,
    Chalmers University of Technology,
    Gothenburg, Sweden%
}
\newcommand{\affilpsi}{%
    Photon Science Division,
    Paul Scherrer Institute,
    Villigen,
    Switzerland%
}
\newcommand{\affilepfl}{%
    Institute of Materials,
    \`Ecole Polytechnique F\`ed\`erale de Lausanne,
    Lausanne,
    Switzerland%
}

\begin{document}

\title{Mumott -- a Python package for the analysis of multi-modal tensor tomography data}

\author{Leonard C. Nielsen}
\affiliation{\chalmersphys}
\author{Mads Carlsen}
\affiliation{\affilpsi}
\author{Sici Wang}
\affiliation{\affilpsi}
\affiliation{\affilepfl}
\author{Arthur Baroni}
\affiliation{\affilpsi}
\author{Torne T\"anzer}
\affiliation{\affilpsi}
\affiliation{\affilepfl}
\author{Marianne Liebi}
\affiliation{\chalmersphys}
\affiliation{\affilpsi}
\affiliation{\affilepfl}
\email{marianne.liebi@psi.ch}
\author{Paul Erhart}
\affiliation{\chalmersphys}
\email{erhart@chalmers.se}

\date{\today}

\begin{abstract}
Small and wide angle x-ray scattering tensor tomography are powerful methods for studying anisotropic nanostructures in a volume-resolved manner, and are becoming increasingly available to users of synchrotron facilities.
The analysis of such experiments requires, however, advanced procedures and algorithms, which creates a barrier for the wider adoption of these techniques.
Here, in response to this challenge, we introduce the \mumott{} package.
It is written in Python with computationally demanding tasks handled via just-in-time compilation using both CPU and GPU resources.
The package is being developed with a focus on usability and extensibility, while achieving a high computational efficiency.
Following a short introduction to the common workflow, we review key features, outline the underlying object-oriented framework, and demonstrate the computational performance.
By developing the \mumott{} package and making it generally available, we hope to lower the threshold for the adoption of tensor tomography and to make these techniques accessible to a larger research community.
\end{abstract}

\maketitle

%\begin{linenumbers}
\section{Introduction}

The properties of numerous materials depend on the hierarchical organization of their basic building blocks ranging from the nanometer to the micrometer scale.
Examples include plant materials assembled from cellulose and lignin \cite{fratzl_hierarchical_2007}, bone constructed of assemblies of mineralized collagen fibrils \cite{reznikov_biomin_2014}, or polymeric materials such as in the structure of semi-crystalline polymers \cite{schrauwen_2004, stribeck_jps_2008, tang_2007}, and liquid-crystalline polymers composed of rigid macromolecules \cite{Gantenbein_3d_2018}.
The study of structure-property relationships of hierarchical materials for applications in biology, the biomedical field, or polymer engineering relies on accurate structural characterization from a wide range of techniques.
X-ray techniques are of particular interest with respect to providing volume-resolved nanostructural information in macroscopic samples.
Due to the high penetration depth and non-destructive nature of the techniques, methods such as x-ray absorption and phase contrast \gls{ct} have played an important role in providing high resolution densimetric measurements of 3D samples \cite{endrizzi2018, ou_2021_xray_imaging}.
In addition to the densimetric fields, the arrangement of nanostructural elements, in particular their direction and degree of alignment, is important for many mechanical and functional properties.
This brings an additional challenge in methodologically bridging between the length scales of the nanostructural building blocks and the macroscopic specimen.
In addition to methods resolving the nanostructure spatially with high-resolution imaging techniques, the orientation of the nanostructure can also be probed by polarization, scattering, diffraction, or magnetic relaxation methods in a volume averaged way \cite{marios_rst_2016}.
The first technique for probing the orientation of structures without directly resolving them was polarized light microscopy.
It was followed by other techniques that also make use of the polarization of light such as polarized Raman or Fourier transform infrared spectroscopy and polarized second-harmonic-generation imaging.
X-ray and neutron diffraction approaches can be used, such as \gls{ddf} imaging\cite{jensen_th_xrayDDF, busi_m_neutronDDF} which probes the orientation at the micrometer-scale through the integrated scattering signal, or scanning small and wide angle scattering that allow probing of nanoscale structures.
\Gls{saxs} probes the spatial variation of the electron density, providing information on microstructural elements with characteristic length scales in the range of tens to hundreds of nanometers, relating to the structural organization and orientation of the materials at the corresponding length scales, while \gls{xrd} (in this paper called \gls{waxs}) probes atomic distances and crystal lattices.
Whereas \gls{ddf} are a family of full-field imaging methods, \gls{saxs} and \gls{waxs} can be used as scanning imaging techniques in which the sample is raster-scanned with a focused x-ray beam providing an image of the sample consisting of a 2D diffraction pattern in each pixel.
Tomographic reconstruction of such measurements using isotropically scattering samples is known as \gls{xrd}-\gls{ct} and is frequently used both in the \gls{saxs} \cite{stribeck_mcp_2006, schroer_apl_2006} and \gls{waxs} \cite{kleuker_xrd_1998, stock_xrdct_2008, bleuet_xrd_2008} regimes at synchrotron x-ray sources.

To access the orientation information of the underlying ultrastructure within a 3D specimen, tomographic methods can be extended from the reconstruction of a scalar fields to tensor fields describing the directionality of the signal, which is in general called \gls{tt}.
The most established technique in this category is diffusion \gls{mri}, also called \gls{dti}, which is widely used to study the 3D arrangement and orientation of neurons.
In the case of X-rays, \gls{tt} has been demonstrated for \gls{ddf} \cite{malecki_ddf_2014, kim_2020_sirt_tt}, \gls{saxs} \cite{liebi_nat_2015, schaff_nature_2015, liebi_aca_2018, gao_aca_2019, nielsen_tt_2023} and \gls{waxs} \cite{grunewald_2020_waxs}.
Other related tomography approaches which can be considered as \gls{tt} include probing magnetic field directions with circular polarized x-rays \cite{donnely_magnetic_3D} or polarized neutrons \cite{sales_2017_threed}.

The acquisition and analysis of \gls{tt} data is a non-trivial undertaking, creating a barrier for the wider adoption of these powerful techniques.
In response to this challenge, specifically with regard to the analysis of such data, we here introduce the software package \mumott{} for the reconstruction of \gls{tt} data.
While the current implementation supports the cases of \gls{saxs} and \gls{waxs}, the framework offers the possibility to include other modalities in the future.
In the following we first provide a brief overview of the methodology (\autoref{sect:methodology}) before describing the structure and functionality of \mumott{} (\autoref{sect:implementation}).
Finally, we give a short outlook concerning potential future additions and developments (\autoref{sect:outlook}).

\section{Methodology}
\label{sect:methodology}

\Gls{saxs}/\gls{waxs}-\gls{tt} is conceptually similar to \gls{xrd}-\gls{ct} in that the sample is raster-scanned though a focused beam to produce a number of 2D projections, varying the sample orientation between each projection.
Unlike \gls{xrd}-\gls{ct} it works with azimuthally re-grouped detector images where the intensity of the scattered x-rays in a number of azimuthal bins is recorded rather than a single azimuthally integrated intensity.
The width of the azimuthal bins depends on the desired angular resolution of the reconstruction.
The azimuthal re-grouping can be done with a number of freely available software tools such as \textsc{pyFAI} \cite{KieValBla20} and \textsc{matfraia} \cite{JenChrWen22}.
The experimental data is then a five-dimensional dataset consisting of the tomographic rotation, the two directions of the raster scan grid, the scattering angle $2\theta$, and the azimuthal angle $\varphi$.
\mumott{} deals with the reconstruction of such a five-dimensional dataset into a six-dimensional reconstruction, consisting of a three-dimensional voxel map containing a three-dimensional \gls{rsm} in each voxel.

We assume that the data has already been corrected for various experimental errors pertaining to solid angle, geometric distortions, and polarization.
To account for the effect of absorption by the sample, the collected data can be normalized by the transmitted intensity as is common practice in \gls{xrd}-\gls{ct}.
Especially at small scattering angles, this makes it possible to carry out reconstructions even with low sample transmission coefficients ($\approx$1\% has been demonstrated) assuming sufficient incident flux.
The measurement of the transmitted beam intensity can be done using either a semi-transparent beam stop, a diode mounted on the beam stop, or a fluorescence measurement \cite{pauw_saxs_review}.
Alternatively, synthetic transmission data can be calculated based on an absorption \gls{ct} reconstruction\cite{grunewald_2023_iucrj}. 

\begin{figure}
\includegraphics[width=0.8\linewidth]{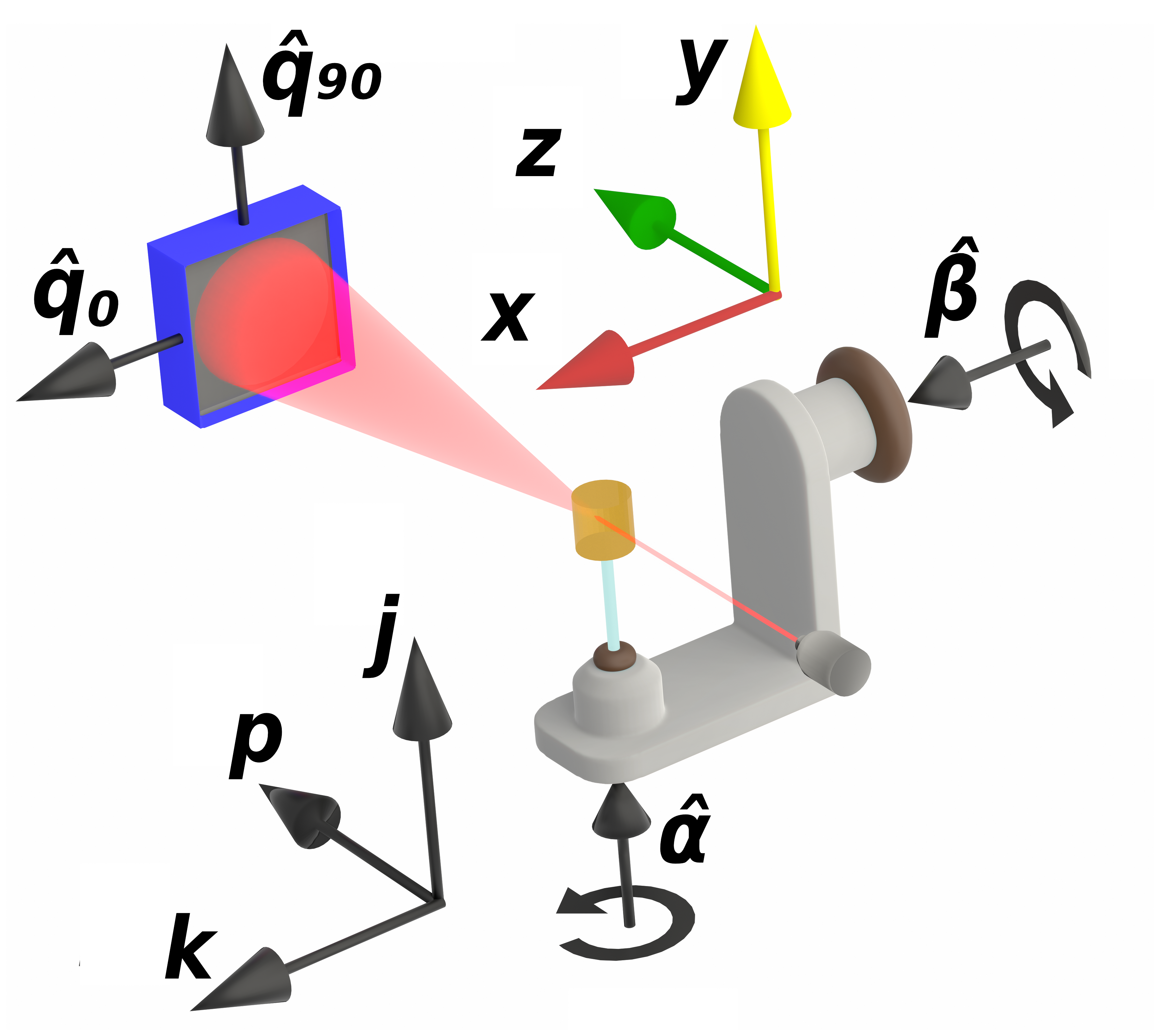}
\caption{
    Illustration of vectors defining the experimental geometry in the laboratory coordinates (i.e., at $\alpha = \beta = 0$).
}
\label{fig:exp_setup}
\end{figure}

The experiment is described in a coordinate system defined by the voxel grid of the sample and the three orthogonal basis vectors $\hat{\vec{x}}$, $\hat{\vec{y}}$, and $\hat{\vec{z}}$.
Typically these vectors are chosen to conform to the convention of the beamline, where the experiments were performed, such that the sample-fixed coordinates correspond to the laboratory coordinates when the goniometer angles are zeroed.
The geometry of the instrument is defined by specifying a number of unit vectors in these laboratory coordinates.
These vectors include the beam direction $\hat{\vec{p}}$ (also called the projection vector), the two orthogonal directions of the raster-scan $\hat{\vec{j}}$ and $\hat{\vec{k}}$, and two vectors describing the origin and the positive direction of the azimuthal integration $\hat{\vec{q}}_0$ and $\hat{\vec{q}}_{90}$ defined by the equation
\begin{equation}
\begin{split}
    \hat{\vec{q}}(\varphi) &= \cos(2\theta/2)(\cos\varphi\hat{\vec{q}}_0 
    + \sin\varphi\hat{\vec{q}}_{90}) - \sin(2\theta/2)\hat{\vec{p}} \\
    &\approx \cos\varphi\hat{\vec{q}}_0 
    + \sin\varphi\hat{\vec{q}}_{90}.
\end{split}
\end{equation}
This equation gives the normalized scattering vector $\hat{\vec{q}}(\varphi)$ probed by each detector segment as a function of the scattering angle $2\theta$ and the detector azimuth angle $\varphi$.
The second line gives a useful approximation valid for small scattering angles.
The sample can be rotated by a goniometer and the rotation of the sample goniometer at a given setting labeled by $s$ results in a rotation matrix $\matrix{R}_s$.
Typically, the goniometer is constructed by two orthogonal rotation stages, an inner ``rotation'' and outer ``tilt'' stage.
The full rotation is then defined by a pair of rotation angles $\alpha$ and $\beta$ with corresponding rotation axes $\hat{\vec{\alpha}}$ and $\hat{\vec{\beta}}$, such that $\matrix{R}_s = \matrix{R}_{\hat{\vec{\beta}}}(\beta)\matrix{R}_{\hat{\vec{\alpha}}}(\alpha)$.
While all these vectors may be chosen freely in \mumott{} (under the restriction that certain vectors are orthogonal to certain other vectors), we work in a standard geometry in this paper given by the choices tabulated in \autoref{tab:geometry_vectors} and visualized in \autoref{fig:exp_setup}.

\begin{table}
    \caption{
        Unit vectors defining the experimental geometry and their values in the standard geometry used in previous publications such as Ref.~\citenum{liebi_aca_2018}.
    }
    \label{tab:geometry_vectors}
    \begin{center}
    \begin{tabular}{*{2}cl}
         \toprule
         Symbol
         & Standard
         & Field name \\
         \midrule
         $\hat{\vec{p}}$       & $+\hat{\vec{z}}$  & \texttt{p\_direction\_0} \\
         $\hat{\vec{j}}$       & $+\hat{\vec{y}}$  & \texttt{j\_direction\_0} \\
         $\hat{\vec{k}}$       & $+\hat{\vec{x}}$  & \texttt{k\_direction\_0} \\
         $\hat{\vec{q}}_0$     & $+\hat{\vec{x}}$  & \texttt{detector\_direction\_origin} \\
         $\hat{\vec{q}}_{90}$  & $+\hat{\vec{y}}$  & \texttt{detector\_direction\_positive\_90} \\
         $\hat{\vec{\alpha}}$  & $+\hat{\vec{y}}$  & \texttt{inner\_axis} \\
         $\hat{\vec{\beta}}$   & $+\hat{\vec{x}}$  & \texttt{outer\_axis} \\
         \bottomrule
    \end{tabular}
    \end{center}
\end{table}

\begin{figure}
    \centering
    \includegraphics[width=0.9\linewidth]{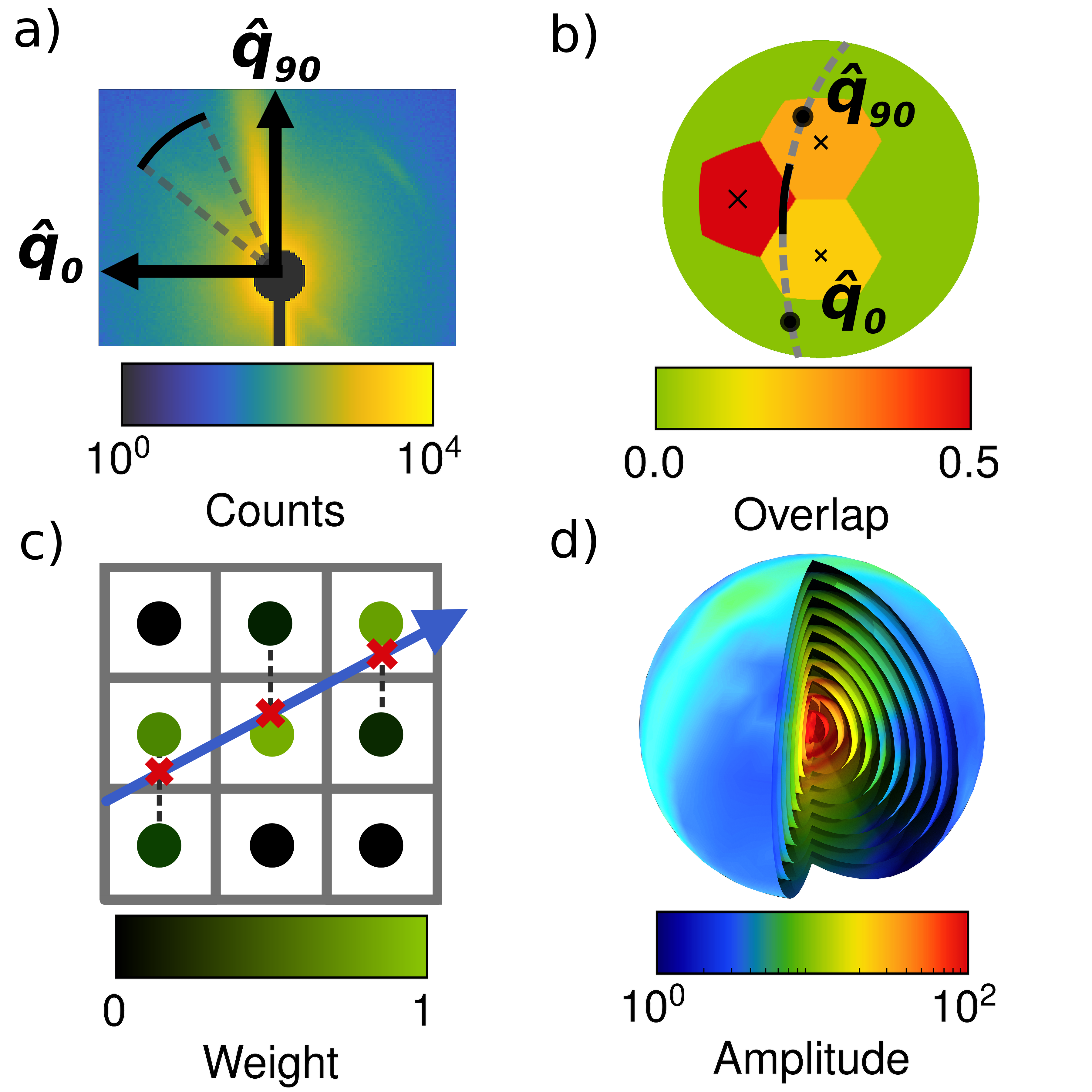}
    \caption{
        (a) Layout of vectors and angles on the detector.
        A single detector segment is marked with a thick black line.
        (b) Integrated basis function values $B_{sc,i}$ plotted in a stereographic projection.
        The black solid arc corresponds to the single detector segment marked in (a).
        (c) Computed probing of each voxel by bilinear interpolation.
        (d) Splitting of a 3D-\gls{rsm} into a stack of 2D-\glspl{rsm} at fixed $q$-lengths.
    }
    \label{fig:methodology}
\end{figure}

The scattering from a given voxel $(x, y,z)$ is proportional to a characteristic function $f^{\mathrm{3D}}_{xyz}(\vec{q})$ called the \textit{3D \acrfull{rsm}}.
In the context of \gls{saxs}-\gls{tt}, the \gls{rsm} is the Fourier transform of the auto-correlation function of the electron density taken over a small volume.
For the purpose of reconstruction, we consider one ``shell'' of reciprocal space at a time and the 3D \gls{rsm} is built up by reconstructing and stacking successive 2D shells (sketched in \autoref{fig:methodology}d).
For one such shell we consider the function $f^{\mathrm{2D}}_{xyz}(\hat{\vec{q}})$, which depends only on the \emph{direction} of the scattering vector.
This function is modeled by a sum of basis functions,
\begin{equation}
    f^{\mathrm{2D}}_{xyz}(\hat{\vec{q}}) = \sum_i c_{xyzi} B_i(\hat{\vec{q}})
    \label{eq:basis_set_expansion},
\end{equation}
where $B_i(\hat{\vec{q}})$ are the basis functions (see \autoref{sect:basis_sets} below) and $c_{xyzi}$ are the unknown expansion coefficients that we want to find.

$f^{\mathrm{2D}}_{xyz}(\hat{\vec{q}})$ is described in sample-fixed coordinates such that at a given rotation of the sample given by $\matrix{R}_s$ the detector segment at the angle $\varphi$ measures the component $f^{\mathrm{2D}}_{xyz}(\matrix{R}_s^{\mathrm{\intercal}}\hat{\vec{q}}(\varphi))$, where the superscript $\{\}^{\mathrm{\intercal}}$ denotes the matrix transpose.
In the normal setting, the detector is split into a number of evenly spaced segments indexed by $c$ covering either the full \qty{360}{\degree} for \gls{waxs} or \qty{180}{\degree} for \gls{saxs}.
Each detector segment (\autoref{fig:methodology}a) is defined by a start angle $\varphi_{c,\mathrm{start}}$ and an end angle $\varphi_{c,\mathrm{end}}$.
As such, the detector segment probes the average of the scattering function within this interval given by the integral
\begin{equation*}
    I_c \propto \frac{1}{\varphi_{c,\mathrm{end}}-\varphi_{c,\mathrm{start}}}\int_{\varphi_{c,\mathrm{start}}}^{\varphi_{c,\mathrm{end}}} f^{\mathrm{2D}}_{xyz}\left(\matrix{R}_s^{\mathrm{\intercal}}\hat{\vec{q}}(\varphi)\right) \mathrm{d}\varphi
\end{equation*}
by inserting Eq.~\eqref{eq:basis_set_expansion} into the above, we define the constants (\autoref{fig:methodology}a,b)
\begin{equation}
\begin{split}
    B_{sc,i} &= \frac{1}{\varphi_{c,\text{end}}-\varphi_{c,\text{start}}}\int_{\varphi_{c,\text{start}}}^{\varphi_{c,\text{end}}} B_i\left(\matrix{R}_s^{\mathrm{\intercal}}\hat{\vec{q}}(\varphi)\right) \mathrm{d}\varphi,
    \label{eq:projection_matrix}
\end{split}
\end{equation}
which describe how much each basis function scatters in the direction measured by a given detector segment, illustrated in \autoref{fig:methodology}a.
This is an integral over a single scalar variable, which can be numerically evaluated by standard methods of quadrature.

To complete the forward model, we have to sum up the intensity contributions from all voxels in the path of the incident beam.
At a given position of the raster scan and rotation of the goniometer, only the voxels that are illuminated by the beam contribute to the measured scattering.
A given voxel is indexed by the three integers $x$, $y$ and $z$ and at a given setting of the sample goniometer the position of the voxel is
\begin{equation}
    \vec{r}_{xyz} = a \matrix{R}_s \begin{bmatrix}
        x &
        y &
        z
    \end{bmatrix}^{\intercal}
    - b (j - \delta j) \hat{\vec{j}}-b (k - \delta k) \hat{\vec{k}},
    \label{eq:real_space_geom}
\end{equation}
where $j$ and $k$ are integer indices of the raster scan, $a$ is the step size of the cubic voxel grid, $b$ is the step size of the 2D raster scan, and $ \delta j$ and $ \delta k$ are offsets caused by parasitic movements of the sample stage during rotation. Typically the resolution of the reconstruction is matched with the raster scan such that $a = b$.

Finally, to include the scattering from all probed voxels, we introduce the factor $P_{sjk,xyz}$ which describes how much the $xyz$-voxel overlaps with the incoming beam at the position $j,k$ of the raster scan at the goniometer setting $s$.
$P_{sjk,xyz}$ takes a value between 0 and 1 with the value 0 for any voxel that does not intersect the x-ray beam.
Using this factor we can now write the scattered intensity as a sum over all voxels in the voxel grid:
\begin{equation}
    I(\varphi)_{sjk} = \sum_{xyz} P_{sjk,xyz}
    f^{\mathrm{2D}}_{xyz}(\matrix{R}_s^{\mathrm{\intercal}}\hat{\vec{q}}(\varphi)).
    \label{eq:projection_operator}
\end{equation}

Figure~\ref{fig:methodology}c gives a graphical interpretation of the $P_{sjk,xyz}$ coefficients.
By combing Eqs.~\eqref{eq:basis_set_expansion} and \eqref{eq:projection_operator} we can now write up the full forward model for \gls{tt}:
\begin{align}
    I_{sjkc} = &\sum_{xyz} P_{sjk,xyz} \sum_i B_{sic} c_{xyzi} \\
    &\Leftrightarrow \matrix{I} = \matrix{A}\vec{c},
    \label{eq:forward_model}
\end{align}
where on the second line we have defined the data vector $\matrix{I}$, the system matrix $\matrix{A}$, and the coefficient vector $\vec{c}$ in order to write the problem in linear algebra terms.
The system matrix has the block matrix structure
\begin{equation}
    \matrix{A} = \begin{bmatrix}
        [P_{0jk,xyz}] \otimes [B_{0c,i}] \\
        [P_{1jk,xyz}] \otimes [B_{1c,i}] \\
        \vdots \\
        [P_{N_sjk,xyz}] \otimes [B_{N_sc,i}] 
    \end{bmatrix},
\end{equation}
where $\otimes$ is the Kronecker product.
Note that the system matrix does not factorize into a projection part and a reciprocal-space part, as both the projection operator and the basis function matrix depend on the orientation of the sample.
This structure highlights the difference between tensor tomography and many other multi-modal tomography techniques such as \gls{xrd}-\gls{ct}, x-ray fluorescence tomography\cite{dejonge_XFT}, time-resolved tomography, and spectral tomography \cite{shikhaliev_spectral_ct}, where the real-space projection operation and the mapping of the other modality are decoupled.
This prevents the use of many techniques that rely on this factorization such as principal-component-analysis methods \cite{gao_sparsity}.

With the forward model defined, we can now formulate the inversion as the solution of a minimization problem:
\begin{equation}
    \vec{c}^* = \\
    \underset{\vec{c}}{\argmin}\left[ \left\lVert \matrix{I} - \matrix{A}\vec{c} \right\rVert^ {a}_{a} \\
    + \mu \left\lVert\matrix{D}\vec{c} \right\rVert^{b}_{b} + \ldots \right], \label{eq:opt_prob_general_form}
\end{equation}
where $\left\lVert\cdot\right\rVert_{a}$ and $\left\lVert\cdot\right\rVert_{b}$ are two, potentially identical, vector norms, $\mu$ is a regularization parameter, and $\mathrm{D}$ is a weight matrix.
The ellipsis indicates that more regularization terms of the same form as $\mu \left\lVert\matrix{D}\vec{c} \right\rVert^{b}_{b}$ may be added.

\section{Implementation}
\label{sect:implementation}

\mumott{} is written in Python with performance-critical parts implemented using the \numba{} package \cite{numba} for CPU and GPU acceleration in order to balance computational efficiency, portability, and maintainability.
It also depends on \textsc{numpy} \cite{numpy}, \textsc{scipy} \cite{scipy}, \textsc{scikit-image} \cite{scikit-image}, and \textsc{colorcet} \cite{colorcet}.
The package is extensively documented and the documentation is available online at \url{https://mumott.org} and at \url{https://doi.org/10.5281/zenodo.7919448}, including various examples in the form of Jupyter notebooks.

A variety of common tasks pertaining to data alignment and reconstruction is accessible via functions that provide a rather simple yet customizable interface.
These functions represent so-called ``pipelines'' (\autoref{sect:pipelines}) and are intended to serve as the primary interface for most users.

The pipeline functions combine a number of individual tasks and components, which are represented via objects and are part of the underlying object-oriented framework (\autoref{sect:oo-framework}).
Through the latter, advanced users and developers can customize, adapt, and extend the functionality of \mumott{}.
\mumott{} is released under the Mozilla Public License Version 2.0 and developed as free-and-open-source software, inviting the contributions of other groups and developers.

In the following, we first provide a short demonstration of the workflow (\autoref{sect:workflow}) before addressing basis sets (\autoref{sect:basis_sets}), several common pipelines (\autoref{sect:pipelines}), the underlying object-oriented framework (\autoref{sect:oo-framework}), and computational efficiency (\autoref{sect:computational-efficiency}).

\subsection{Workflow}
\label{sect:workflow}

Tables~\ref{scheme:example_workflow_1} and \ref{scheme:example_workflow_2} show examples of simple workflows in \mumott{} for reconstructing a voxel map of 2D-\glspl{rsm} from experimental data.
In the following sections we explain each of the steps in this process.

\begin{table}
\caption{
    Minimal example of a reconstruction workflow using the \acrfull{mitra} pipeline (\autoref{sect:standard-pipelines}).
}
\label{scheme:example_workflow_1}
\begin{lstlisting}[style=Python]
# Load data
data_container = DataContainer('trabecular_bone.h5')

# Perform alignment
shifts, _, _ = run_optical_flow_alignment(
    data_container, use_gpu=True)
data_container.geometry.j_offsets = shifts[:, 0]
data_container.geometry.k_offsets = shifts[:, 1]

# Execute reconstruction pipeline
result = run_mitra(data_container)   
\end{lstlisting}
\end{table}

\begin{table}
\caption{
    Extended example for a reconstruction workflow using a Tikhonov ($L_2$) regularized least-squared model and spherical harmonics as basis functions that uses the object-oriented interface (\autoref{sect:oo-framework}).
}
\label{scheme:example_workflow_2}

\begin{lstlisting}[style=Python,mathescape=true]
# Load data
data_container = DataContainer('trabecular_bone.h5')

# Perform alignment
shifts, _, _ = run_optical_flow_alignment(
    data_container, use_gpu=True)
data_container.geometry.j_offsets = shifts[:, 0]
data_container.geometry.k_offsets = shifts[:, 1]

# Define forward model
projector = SAXSProjectorCUDA(
    data_container.geometry)
basis_set = SphericalHarmonics(ell_max=8)
residual_calculator = GradientResidualCalculator(
    data_container, basis_set, projector)
loss_function = SquaredLoss(residual_calculator)
l2_norm = L2Norm()
loss_function.add_regularizer(
    name='l2norm', regularizer=l2_norm,
    regularization_weight=2e-6)

# Carry out reconstruction
optimizer = LBFGS(loss_function, maxiter=20)
result = optimizer.optimize()    
\end{lstlisting}
\end{table}

\subsubsection{Loading the data.}
The acquired data (after beamline specific preprocessing) is handled using a \texttt{DataContainer}, which is created by loading a HDF5 file (\autoref{tab:h5_format}) that contains the azimuthally re-grouped data for one $q$-bin of the experiment.
While a full experimental dataset containing a detector frame for each scan position can be quite large, commonly on the order of 100s of gigabytes, a single $q$-bin of the azimuthally regrouped data is usually hundreds of megabytes to a few gigabytes.
The data files can be prepared containing the geometry data and sample offsets information or only the data.

\subsubsection{Definition of the geometry.}
The geometry is defined by the vectors listed in \autoref{tab:geometry_vectors}, which can be given in any consistent coordinate system.
In the examples shown, the full geometry data is already contained in the data file (and hence the \texttt{DataContainer}), but in general it is possible to override certain parameters after loading.

\subsubsection{Aligning the data.}
Before a meaningful reconstruction can be carried out the data must be aligned, which means to calculate the offsets defined in Eq. \eqref{eq:real_space_geom}.
To this end, \mumott{} provides several pipelines that use the transmission measurement or the average scattering to correct misalignment between each projection that occur due to parasitic movements during acquisition.
In the examples shown here, we use the function that implements the optical flow alignment procedure \cite{Odstrcil:19}, which relies on center-of-mass and tomographic consistency techniques.
The alignment functions return most importantly the shifts that are needed for aligning the data.
These values are then used to override the offsets stored in the \texttt{DataContainer} object.

\subsubsection{Defining the reconstruction model.}
The reconstruction model is defined by the choice of basis functions, the form of the cost function as well as the regularization terms.
A large number of different algorithms can be constructed by combining these three choices.
The simplest approach is to utilize one of the existing pipelines (\autoref{sect:pipelines}) as illustrated by the first example (\autoref{scheme:example_workflow_1}), in which the \gls{mitra} pipeline (\autoref{sect:standard-pipelines}) is used.
Alternatively one can configure a reconstruction model using the individual objects that represent the different components.
This approach is demonstrated by the second example (\autoref{scheme:example_workflow_2}), where we choose a basis of spherical harmonics in combination with a squared-difference loss function and Tikhonov ($L_2$) regularization.

\subsubsection{Minimizing the loss function.}
Once the loss function is defined, the optimization problem can be solved using one of a number of optimization routines.
While this step is included in the case of the predefined reconstruction pipeline in the first example (\autoref{scheme:example_workflow_1}), it needs to be explicitly specified when constructing the workflow as in the second example (\autoref{scheme:example_workflow_2}), where we use the gradient-based LBFGS optimizer.
In the case of regularized models, one should then perform a sweep of the regularization parameter space in order to determine (a) sensible regularization parameter(s).

\begin{table}
\caption{
    Outline of the HDF5 file format used by \mumott{}.
    The spaces indicate the hierarchy of entries; \texttt{0} is an entry in the group \texttt{projections}, whereas \texttt{data} is an entry in the group \texttt{0}, and so on.
}
\label{tab:h5_format}
\begin{tabular}{*{3}ll}
     \toprule
     Path & Type \\
     \midrule
     \texttt{p\_direction\_0}  & $\mathtt{float}(3)$\\
     \texttt{j\_direction\_0}  & $\mathtt{float}(3)$\\
     \texttt{k\_direction\_0}  & $\mathtt{float}(3)$\\
     \texttt{detector\_direction\_origin}  & $\mathtt{float}(3)$\\
     \texttt{detector\_direction\_positive\_90}  & $\mathtt{float}(3)$\\
     \texttt{inner\_axis}  & $\mathtt{float}(3)$\\
     \texttt{outer\_axis}  & $\mathtt{float}(3)$\\
     \texttt{volume\_shape}  & $\mathtt{int}(3)$\\
     \texttt{detector\_angles}  & $\mathtt{float}(n_\varphi)$\\
     \texttt{projections}  & Group\\
      \quad \quad \texttt{0} &  Group\\
      \quad \quad \quad \quad\texttt{data} & $\mathtt{float}(n_j, n_k, n_\varphi)$\\
      \quad \quad \quad \quad\texttt{diode} & $\mathtt{float}(n_j, n_k)$\\
      \quad \quad \quad \quad\texttt{inner\_angle} & $\mathtt{float}(1)$\\
      \quad \quad \quad \quad\texttt{j\_offset} & $\mathtt{float}(1)$\\
      \quad \quad \quad \quad\texttt{k\_offset} & $\mathtt{float}(1)$\\
      \quad \quad \quad \quad\texttt{outer\_angle} & $\mathtt{float}(1)$\\
      \quad \quad \quad \quad\texttt{weights} & $\mathtt{float}(n_j, n_k, n_\varphi)$\\
      \quad \quad \texttt{1} &  Group\\
      \quad \quad  \vdots  & \\
     \bottomrule \\
\end{tabular}
\end{table}

\subsection{Deriving standard quantities from the output.}

The result of a tensor tomography reconstruction is the array of optimized coefficients, $\mathbf{c}^* = [c^*_{xyzi}]$, which are the voxel-by-voxel expansion coefficients of the local 2D-\gls{rsm} shells in terms of the specific basis functions.
In general, the coefficients can be interpreted using the corresponding basis set to compute latitude-longitude maps of the 2D-\glspl{rsm} shells and reconstructions of several q-bins can be combined to construct 3D-\glspl{rsm} from these maps. 
Furthermore, a number of derived quantities are conventionally used for evaluation and visualization of reconstructions and can be calculated efficiently from the coefficients without needing to compute latitude-longitude maps.
Note that we define here the derived quantities which are part of the output structure of \mumott{} VX.X, additional quantities can be calculated from the array of optimized coefficient, depending on the basis function.

The mean scattering intensity, also called the isotropic intensity, is a measure of the density of scattering material in all orientations and is defined as:

\begin{equation}
\begin{split}
    \overline{f} &= \left\langle f^{\mathrm{2D}}_{xyz}(\hat{\vec{q}})\right\rangle_{\hat{\vec{q}}} \\
    &= \frac{1}{4\pi} \int_0^{\pi}\left(\int_0^{2\pi} f^{\mathrm{2D}}_{xyz}(\hat{\vec{q}}(\theta, \phi))\mathrm{d}\phi\right)\sin\theta\mathrm{d}\theta,
\end{split}
\end{equation}

where $\theta$ and $\phi$ are a pair of polar coordinates for the unit sphere.

A 2D-\glspl{rsm} can also be expanded in tensor-components. Especially the rank-2 tensor is of interest because it allows easily computing primary directions, given by the eigenvectors of the matrix.
The second moment tensor is a 3 by 3 matrix with elements:
\begin{equation}
    M_{ij} = \left\langle q_iq_j f^{\mathrm{2D}}_{xyz}(\hat{\vec{q}})\right\rangle_{\hat{\vec{q}}},
\end{equation}
where $q_i$ are the $x$, $y$, and $z$ components of $\hat{\vec{q}}$ for $i = 1, 2, 3$ respectively.

\begin{figure}
    \centering
    \includegraphics[width=\linewidth]{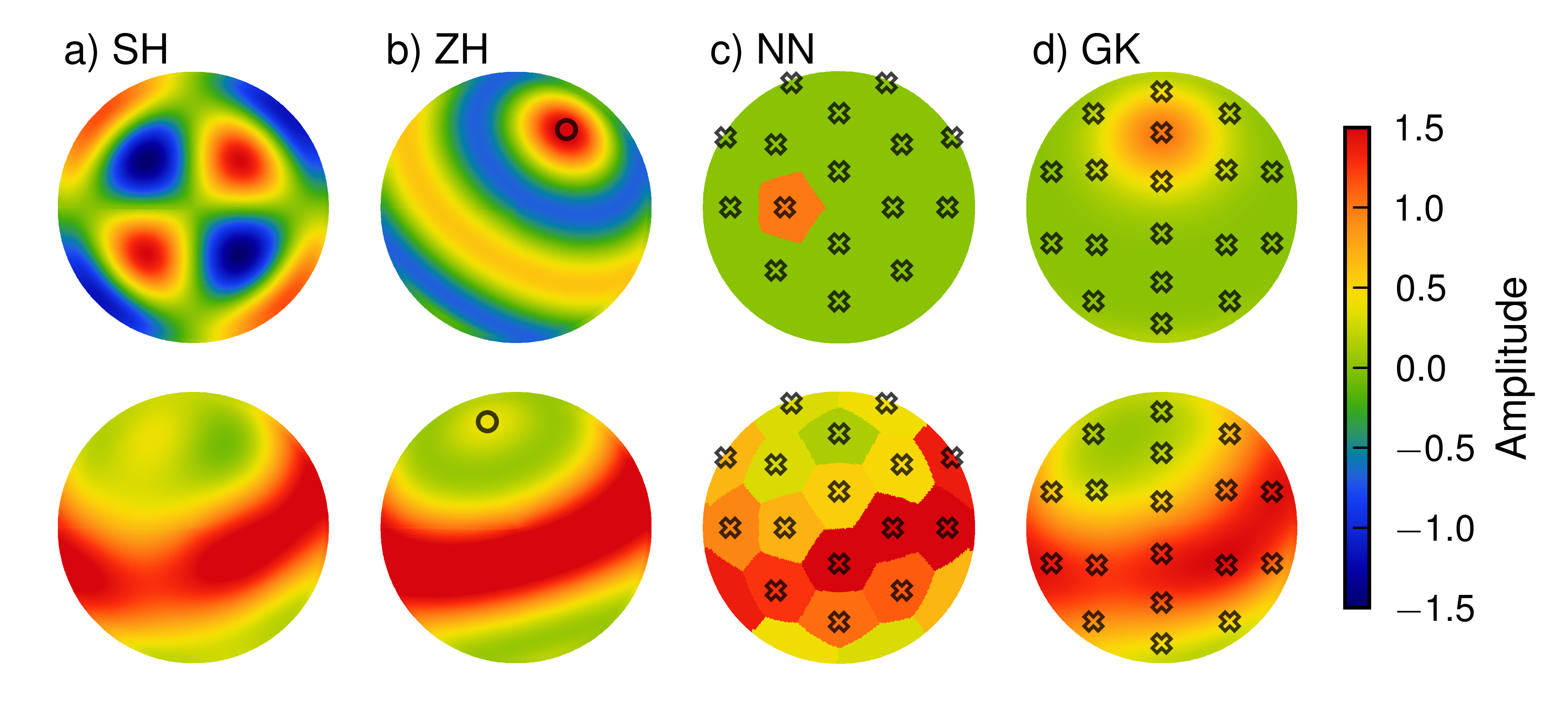}
    \caption{
        Comparison of (a) \acrfull{sh}, (b) \acrfull{zh}, (c) \acrfull{nn}, and (d) \acrfull{gk} basis sets.
        The upper row shows a single basis function for each basis set.
        The lower row shows the \gls{rsm} at a single $q$ of a single voxel of a reconstruction using each of the four basis sets.
        The crosses in (c--d) show the grids which are part of the definition of the \acrshort{nn} and \acrshort{gk} basis sets.
        The directions used to define the \acrshort{nn} model are the face centers of the truncated icosahedron.
        The directions used in the \acrshort{gk} model are given by a modified Kurihara mesh.
        The circles in (b) indicate the symmetry axes.
    }
    \label{fig:basis_sets}
\end{figure}

For many samples, a main orientation, such as a fiber-symmetry axis of the nanostructure, can be defined per voxel.
The rank-2 tensor provides a means to efficiently compute this direction through its eigendecomposition.
However, we point out that the interpretation of the main orientation of the nanostructure depends on its scattering characteristics. Structures where a single direction of strong scattering is expected at two opposite poles, the main orientation is the eigenvector corresponding to the largest eigenvalue.
Similarly, for samples where a ring equatorial band with strong scattering is observed (e.g., the structure displayed in \autoref{fig:basis_sets}) the eigenvector corresponding to the smallest eigenvalue should be chosen.
This provides a fast and noise-tolerant approach to finding the main nanostructure orientation except in cases where the rank-2 term vanishes, such as in Bragg scattering from cubic-symmetric materials, where more advanced approaches are needed.

Another quantity frequently used to describe the anisotropy \cite{BASSER1996209} of tensor tomography is the \gls{fa} which can be computed from the eigenvalues of the 2nd moment tensor:
\begin{equation}
    \mathrm{FA} = \frac{\sqrt{(\lambda_1 - \lambda_2)^2 +(\lambda_2 - \lambda_3)^2 +(\lambda_3 - \lambda_1)^2}}{\sqrt{2(\lambda_1^2+\lambda_2^2+\lambda_3^2)}},
\end{equation}
where $\lambda_1$, $\lambda_2$, $\lambda_3$ are the three eigenvalues of the second moment tensor.
$\mathrm{FA}=0$ for perfectly isotropic scattering and reaches a maximum value of 1 when there is strong scattering in one direction and the scattering goes to zero in the orthogonal directions.
Other values can be calculated from the coefficients to describe the anisotropy, also referred to as the degree of orientation, and are described elsewhere \cite{liebi_nat_2015, nielsen_tt_2023,nielsen_jsr_2024}.  

\subsection{Basis sets}
\label{sect:basis_sets}

The most notable difference between different reconstruction algorithms is the choice of basis functions.
Figure~\ref{fig:basis_sets} shows a comparison of the optimized 2D-\gls{rsm} shell of a single voxel of the same sample using four different basis set types.

\subsubsection{\texorpdfstring{\Gls{sh}}{Spherical harmonics (SH)}.}
The spherical harmonics are a set of orthogonal polynomials that derive from the solution to the Laplace equation in spherical coordinates.
Any function on the unit sphere can be represented by an infinite expansion in spherical harmonics, but in practice the expansion must be truncated at some finite order.
Such a finite expansion in spherical harmonics is called a band-limited spherical function, and can be used to represent the 2D-\gls{rsm} \cite{nielsen_tt_2023}.
The \gls{sh} basis set is fully defined by the band limit, $\ell_{\mathrm{max}}$, at which the expansion in spherical harmonics is truncated.
This sets the resolution of the narrowest diffraction features that can be reconstructed to approximately $2\pi/\ell_{\mathrm{max}}$ radians.

\subsubsection{\texorpdfstring{\Gls{nn}}{Nearest neighbor (NN)}.}
A \gls{nn} basis set uses a set of \gls{nn} indicator functions.
This model is therefore defined by a grid of orientations alone and the resolution is set by the distance between grid points.
This basis set can be used to emulate the algorithm first presented in Ref.~\citenum{schaff_nature_2015}, which splits the \gls{tt} problem into a set of independent scalar tomography problems.

\subsubsection{\texorpdfstring{\Gls{gk}}{Gaussian Kernels (GK)}.}
Like the \gls{nn} basis set, the \gls{gk} basis set is defined by a grid of orientations, but instead of indicator functions it uses smooth spherical Gaussian functions, rotated to be centered on the various grid orientations.
It therefore needs one extra parameter to define the basis set namely the width of the kernel.
The \gls{gk} basis set has many of the same properties as the \gls{nn} basis set but unlike the former results in smooth \glspl{rsm}.
The resolution depends both on the distance between grid points and the kernel width.
Such spherical kernels are commonly used in texture analysis where the specific function used here is referred to as the Bunge normal distribution\cite{Bunge_1969} to distinguish it from several other Gaussian shaped kernel functions that are frequently used.

\subsubsection{\texorpdfstring{\Gls{zh}}{Zonal Harmonics (ZH)}.}
\label{sect:zh-basis-set}

The axially symmetric method established in Ref.~\citenum{liebi_nat_2015} uses a \gls{zh} basis set and thereby differs from the other methods implemented in \mumott{} by having a non-linear forward model.
This requires a separate workflow involving a specialized calculator for the gradients and optimizer.
To enable high-order expansions, simplify the code, and ensure interoperability between the \gls{zh} and \gls{sh} workflows, rotations and gradients are calculated in coefficient space using Wigner D-matrices.
This allows orders up to $\ell_{\mathrm{max}}=100$ in the current implementation, although orders higher than $\ell_{\mathrm{max}}\approx30$ are difficult to handle in practice due of the large number of coefficients.
Details of the implementation are given in \cite{carlsen2024}.

The non-linearity of the forward model in the \gls{zh} approach makes the loss function non-convex, which renders the optimization problem more challenging.
Approaches to overcome this difficulty include regularization of the angle parameters and smoothing of the gradient \cite{liebi_aca_2018}, and the use of an ensemble of randomized starting points \cite{nielsen_tt_2023}.
In \mumott{} we use a starting guess provided by a different reconstruction algorithm to determine the symmetry direction.

\subsection{Pipelines}
\label{sect:pipelines}

\mumott{} provides various pipelines that implement reconstruction and alignment workflows.
The former include both ``standard'' and asynchronous pipelines.
The \emph{standard pipelines} can be run using both CPU and GPU resources and are usually highly customizable.
The \emph{asynchronous} pipelines are optimized for GPU resources and thus speed, and usually slightly less adjustable.
They employ asynchronous execution on the GPU to avoid the overhead caused by transferring data between CPU and GPU.

\subsubsection{Standard reconstruction pipelines.}
\label{sect:standard-pipelines}

The \acrfull{sirt} is a popular reconstruction algorithm thanks to its inherent regularizing properties that result from semi-convergence \cite{elfving_hansen_sirt} and the small number of tunable parameters.
It has previously been used for tensor tomography by, e.g., Ref.~\citenum{schaff_nature_2015} and Ref.~\citenum{kim_2020_sirt_tt}.
In \mumott{} a traditional approach to \gls{sirt}, is implemented in the \gls{sirt} pipeline.

While the \gls{sirt} algorithm is not conventionally stated as a minimization problem, it has been shown that it is equivalent to a specific preconditioned gradient-descent weighted least-squares optimization \cite{gregor_2015_ieee}.
Through this re-formulation, the basic \gls{sirt} reconstruction becomes compatible with various regularizers.
The weight-preconditioner approach employed in this form of \gls{sirt} can also be extended to the \gls{rsm} given by Eq.~\eqref{eq:projection_matrix}.
This approach is implemented in the \acrfull{mitra} pipeline, which permits arbitrary regularizers and basis sets to be used, as well as Nesterov momentum acceleration.

The \acrfull{sigtt} pipeline sets up the basic reconstruction model using a \gls{sh} basis set, a squared-difference loss function, and regularization using a finite-difference Laplacian filter \cite{nielsen_tt_2023}.
The optimization problem is solved with the LBFGS-B algorithms and uses a stop criterion based on the relative change of the loss function.

The \acrfull{dd} pipeline emulates the reconstruction technique used by Ref.~\citenum{schaff_nature_2015}, which splits the tensor reconstruction into a set of independent scalar reconstructions, using the \gls{nn} basis set.
The pipeline employs the \gls{sirt} algorithm for the individual scalar reconstructions.
\Gls{dd} has the practical advantage of needing less VRAM than methods which reconstruct the entire \gls{rsm} at once, as it only loads one scalar component onto the GPU at a time.

\subsubsection{Asynchronous pipelines.}
These pipelines, optimized for GPU execution and speed, include a tensor \gls{sirt} pipeline, which is similar to \gls{mitra} without Nesterov momentum.
In addition, there is \acrfull{motr}, which is essentially the default \gls{mitra} pipeline with $L_1$ and two-sided total variation regularization.
Finally, \acrfull{radtt} optimizes for the Huber norm with two-sided total variation regularization through Nesterov accelerated gradient descent.
The latter pipeline requires fine-tuning of the configuration in order to converge reasonably well, but once a proper step size and smoothing terms are found, it is relatively robust against noise.

There are also sparse versions of the asynchronous pipelines, which use a modified version of the John transform that calculates the reciprocal-space and real-space projection operations simultaneously within one kernel, using a sparse approximation to the reciprocal-space mapping.
This is not necessarily faster than computing the two mappings separately (unless the representation is very sparse, such as only mapping one basis function to each segment), but it uses less VRAM, as it is not necessary to store the intermediate result between carrying out the John transform and carrying out the reciprocal space mapping.

\subsubsection{Alignment.}
\label{sect:alignment-pipelines}

The objective of alignment is to determine the offsets $\Delta j$ and $\Delta k$ of Eq. \eqref{eq:real_space_geom} that result from parasitic movement and misalignment of the goniometer and drift during the experiment \cite{frank_1992_alignment}. 
The alignment step is essential in tomography as any misalignment will be reflected as an artifact in the tomographic reconstruction.
There are many algorithms to solve this problem, leading to sub-pixel alignment accuracy, taking into account various experimental systems and data.

\begin{figure}
    \centering
    \includegraphics[width=\linewidth]{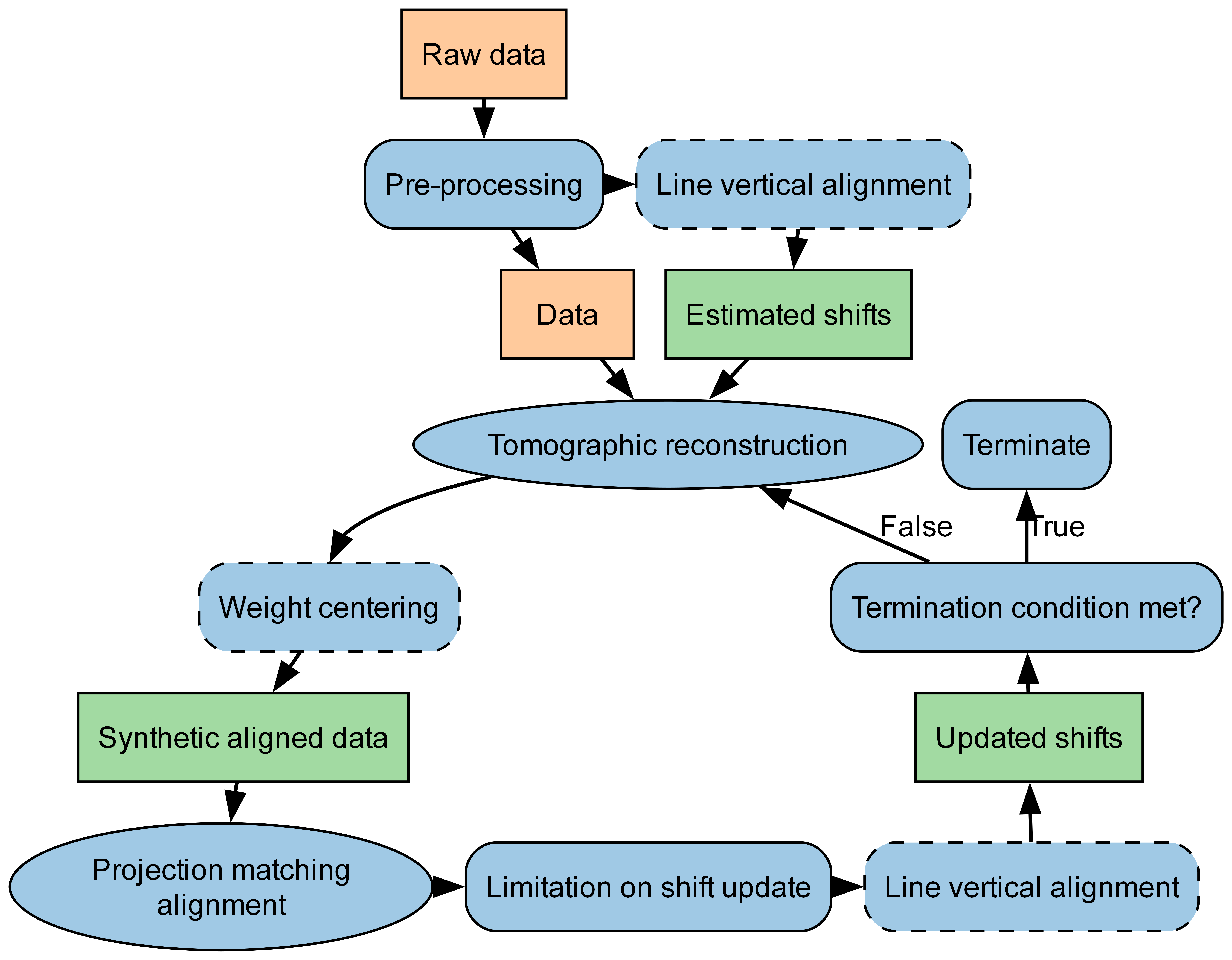}
    \caption{
        Alignment pipeline workflow.
        Steps shown with dashed outlines apply only for the optical flow alignment pipeline.
    }
    \label{fig:alignment_workflow}
\end{figure}

\mumott{} currently provides two alignment pipelines.
Both algorithms typically work with the transmitted intensity data stored in the \texttt{DataContainer} or another isotropic signal such as the azimuthally integrated intensity.
They iteratively update the offsets for each projection by reconstructing the absorption tomogram via a projector.
The overall workflow is shown in \autoref{fig:alignment_workflow}.

The \emph{phase matching alignment} pipeline is based on cross-correlation and follows Ref.~\citenum{manuel_subpixel}.
Cross-correlation alignment has been proven for continuous objects in electron microscopy tomography by Ref.~\citenum{GUCKENBERGER1982167} and has been widely used since.
The principle is to determine the offsets by means of correlation functions formed from image pairs of the projections, comparing the center of mass of the image pair correlation peaks.
This method is fast and can provide sub-pixel accuracy for data with small misalignment.

\begin{figure}
    \centering
    \includegraphics[width=\columnwidth]{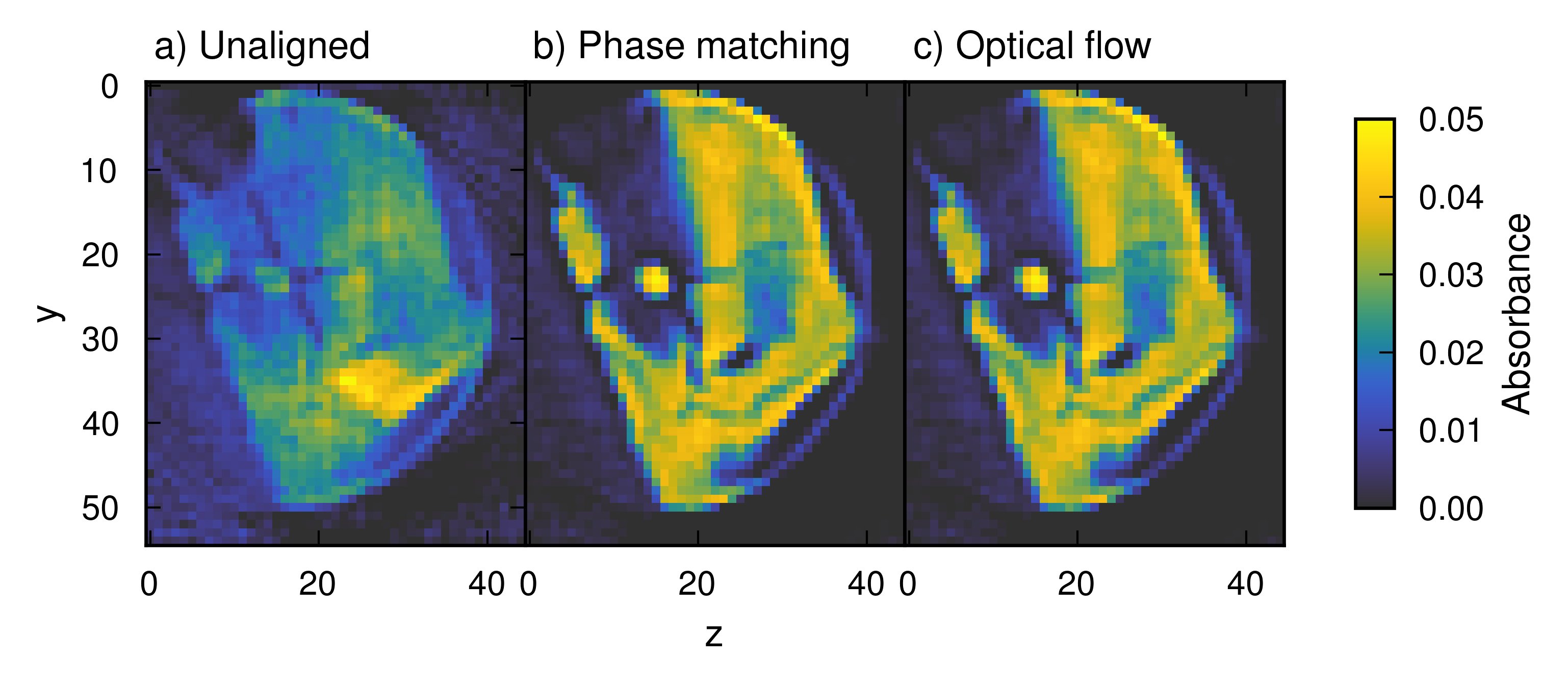}
    \includegraphics[width=\columnwidth]{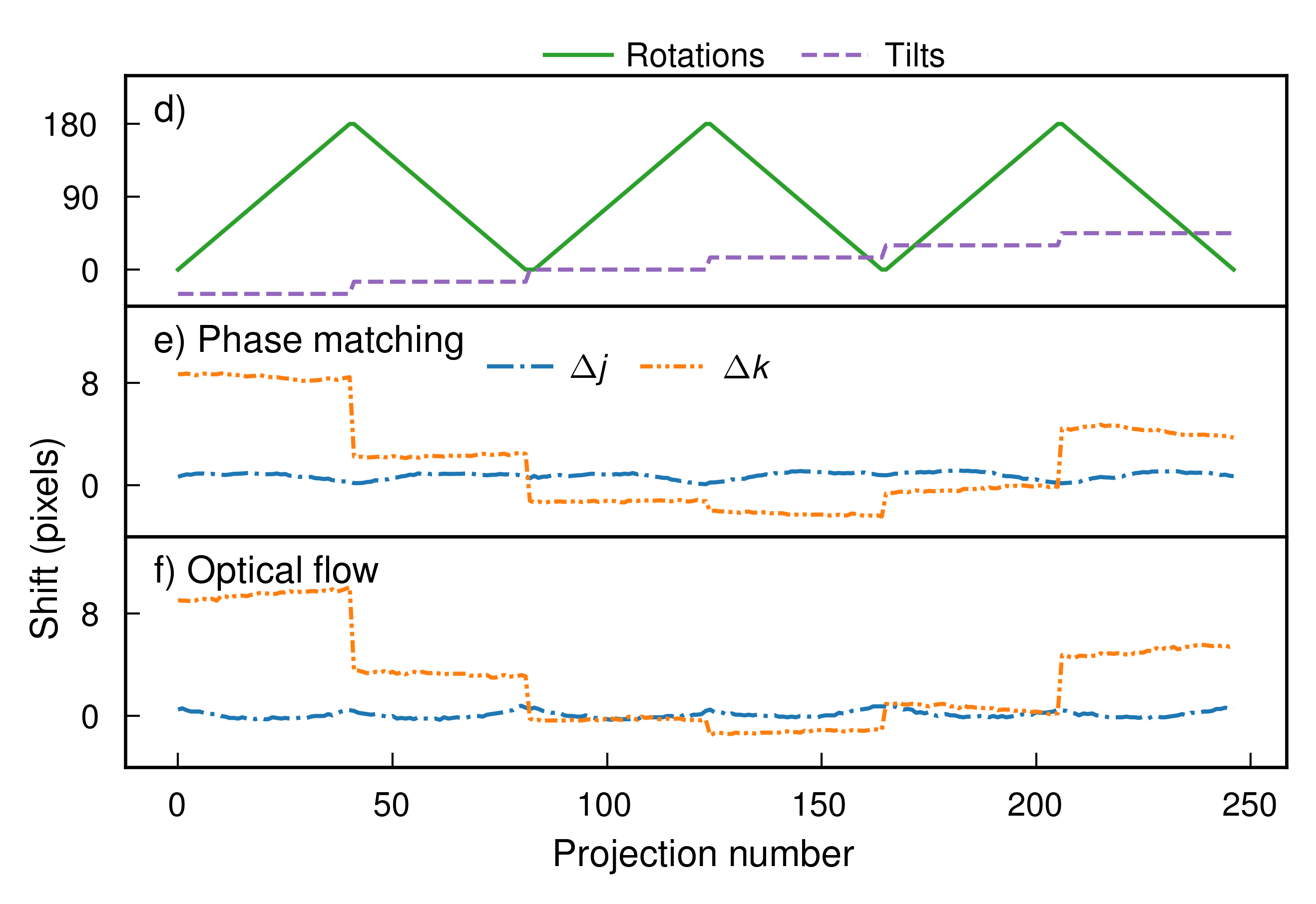}
    \caption{
        Slices from absorption reconstructions (a) before and after alignment with (b) the phase matching method and (c) the optical flow method.
        The projections have been sorted so that the projection directions of neighboring projections are as close to each other as possible.
        (d) Rotations and tilts as well as alignment offsets from (e) the phase matching method and (f) the optical flow method.
        Note how the rotations correlate with changes in $\Delta j$, whereas the tilts correlate with changes in $\Delta k$.
        In this case the offsets result from misalignment of the goniometer's rotation axes with the sample center.
    }
    \label{fig:alignment}
\end{figure}

When the data exhibit misalignment of multiple pixels, the cross-correlation alignment alone can struggle to find the appropriate coordinate transformation.
For such cases, \mumott{} provides the \emph{optical flow alignment} pipeline, which implements a toolbox algorithm based on the work of Ref.~\citenum{Odstrcil:19}. 
This approach uses multiple successive and interconnected alignment procedures, including optical flow projection matching alignment, line vertical alignment, and weight centering.
The method is tunable through various parameters and filters and is therefore able to align extremely misaligned data, providing an approach that is usable for a larger variety of experimental data. 

Examples of alignment results with the two pipelines are shown in the case of the publicly available experimental dataset of trabecular bone in \autoref{fig:alignment} \cite{trabecular_bone_data_set}.

\subsection{Object-oriented framework}
\label{sect:oo-framework}

\begin{figure}
    \centering
    \includegraphics[width=\linewidth]{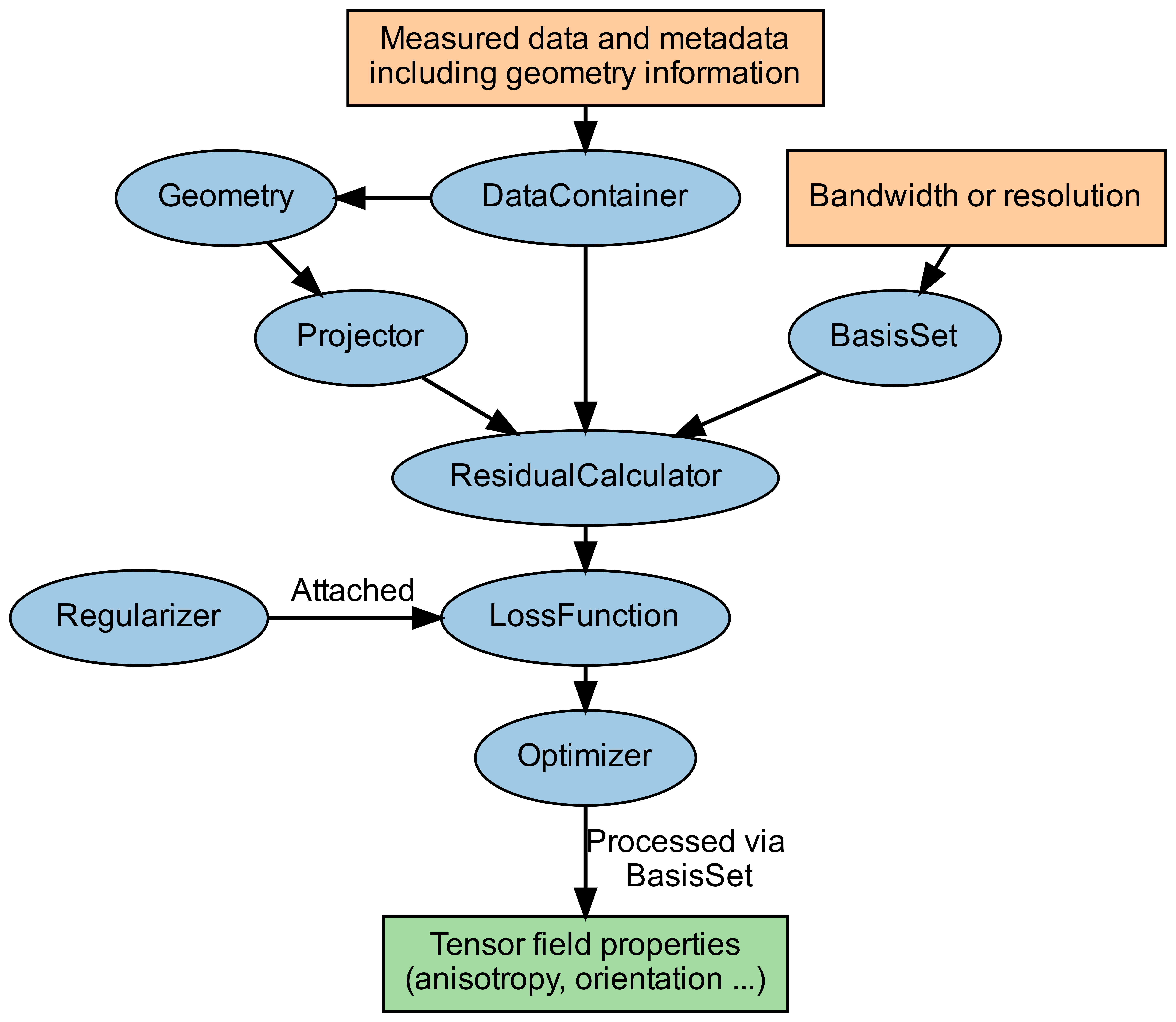}
    \caption{
        Outline of the object-oriented framework in \mumott{}.
        Orange boxes show input parameters and data provided by the user, blue ovals show objects, the green box shows the output, and arrows indicate instances of objects interacting with one another.
    }
    \label{fig:obj_ori_workflow}
\end{figure}

The internal architecture of \mumott{} consists of an object-oriented framework with some elements of functional programming.
The structure of the framework is described in \autoref{fig:obj_ori_workflow}.
Many objects are safely mutable after instantiation and employ hashes of their mutable properties to track the state of linked instances, which means that derived properties can be recomputed when required.

\subsubsection{Data and geometry.}
The \texttt{DataContainer} is the owner of the input data, which is stored in HDF5 format.
The input (measurements and geometry metadata) is stored as a list of projections, indexed by the direction index $s$ as given in Eq.~\eqref{eq:projection_operator}, and the measured tensor tomographic data can be accessed as a four-dimensional array indexed by $[s, j, k, i]$ as in Eq.~\eqref{eq:forward_model}.
The information related to geometry is stored in a \texttt{Geometry} object, which is directly linked to the list of projections attached to \texttt{DataContainer}.
Thus, if a projection is removed from the list, this will be reflected in the corresponding geometry data being removed from the \texttt{Geometry} instance.
The \texttt{Geometry} object stores the basis vectors of the system, i.e., $(\bm{p}, \bm{j}, \bm{k}, \bm{q}_{0}, \bm{q}_{90}, \hat{\bm{\alpha}}, \hat{\bm{\beta}})$ listed in Table~\ref{tab:geometry_vectors} and shown in \autoref{fig:exp_setup}.
The vectors must be specified in the laboratory coordinate system, which coincides with the sample-fixed coordinates $(x, y, z)$ when $\matrix{R}(s) = \mathbf{I}$, \emph{i.e.}, the identity transform.
A rotation operator is then specified, $\bm{R}(s)$, which may be given as a rotation matrix or as an axis-angle quadruplet $(\hat{\bm{\alpha}}, \alpha(s), \hat{\bm{\beta}}, \beta(s))$.
Using the rotation operator, vectors in the sample-fixed coordinates are then dynamically computed for each $s$.
This information can be specified in the input data file or by the user through direct modification of the \texttt{Geometry} object (see, e.g., \autoref{scheme:example_workflow_1}).

\subsubsection{Projectors and basis sets.}
The \texttt{Projector} and \texttt{BasisSet} classes contain the methods and properties needed to compute the forward model defined in Eq.~\eqref{eq:forward_model} and its adjoint.
The \texttt{Projector} objects depends on a \texttt{Geometry} object and employ routines implemented using the \numba{} package \cite{numba} to compute the spatial part of the transform, i.e., the matrix elements $P_{sjk,xyz}$ in Eq.~\eqref{eq:projection_operator}.
This is implemented for both CPU and GPU-based computation, the latter using the \numba{} interface for CUDA.
The implementation employs an approach based on Joseph's method \cite{joseph_jt_ieee} using bilinear interpolation of the field for the forward and the projection for the adjoint computation, respectively, based on the work of Ref.~\citenum{xu_jt_jsb} and Ref.~\citenum{palenstijn_jt_jsb}.

The \texttt{BasisSet} evaluates the constants $B_{sic}$ in Eq.~\eqref{eq:projection_matrix} for the respective basis $\bm{B}$ (\autoref{sect:basis_sets}) using the provided detector geometry and rotation operator $\bm{R}_s^T$.
The integral is evaluated using adaptive Newton-Cotes quadrature or approximated using the central angle of each segment.
In addition, \texttt{BasisSet} provides a routine for computing various properties of reconstructed tensors, such as the orientation as defined by the rank-2 tensor component of the field, the spherical mean, the variance, and relative anisotropy (the spherical standard deviation normalized by the mean).

\subsubsection{Residual calculation.}
The \texttt{ResidualCalculator} is a managing object which takes a \texttt{DataContainer}, \texttt{Projector}, and \texttt{BasisSet}, and uses them to compute residuals.
It tracks the current reconstruction, i.e., $c_{xyzi}$.
In other words, for the data $D_{sjkc}$, and the current reconstruction $c_{xyzi}$ it computes
\begin{equation}
    \mathbf{r} = \mathbf{A}\mathbf{c} - \mathbf{I}
    \label{eq:residual}
\end{equation}
where $\mathbf{r}$ and $\mathbf{I}$ are flattened vectors of the residual and data matrices respectively using the notation introduced in Eq.~\eqref{eq:forward_model}.
It also computes the gradient of a residual norm, which is used by the gradient-based optimization algorithms implemented in \mumott{}.

A special \texttt{ZonalHarmonicsGradientCalculator} is defined to be used as a part of the \gls{zh} workflow.
It is used to map a list of \gls{zh} coefficients and two angle coordinates into the space of all spherical harmonics (up to a maximum order) in the sample-coordinate system, and to compute gradients with respect to the \gls{zh} coefficients and the angles.

\subsubsection{Optimization.}
%While the gradient calculation of the \texttt{ResidualCalculator} can be used directly to implement a gradient-based solver, several common optimization procedures are already implemented.
The goal of the optimization is to minimize a \texttt{LossFunction} (also known as an objective function) by tuning the coefficients of the underlying model by using an \texttt{Optimizer}.
The \texttt{LossFunction} combines a \texttt{ResidualCalculator} with one or several \texttt{Regularizer} instances and can be given a preconditioner to weight the gradient.

There are currently two types of loss functions that support standard least-squares regression (\texttt{SquaredLoss}) and robust regression via the Huber-regressor \cite{huber_1964} (\texttt{HuberLoss}), respectively.

There are also various regularization options.
One can, e.g., smoothen the solution by minimizing the squared $L_2$ norm of the finite-difference Laplacian operator of the tensor field (\texttt{Laplacian}).
It is also possible to smoothen the solution in a more robust manner by minimizing the Huber norm of the spatial gradient for each basis-set mode (\texttt{TotalVariation}).
While it can be more difficult to obtain convergence with more robust terms, it can also be configured to use the Huber approximation for small values to improve convergence.

Other \texttt{Regularizer} classes are available to minimize the $L_2$ and $L_1$ norms of the tensor field, respectively.
While the $L_2$ norm (\texttt{L2Norm}) penalizes large values, which promotes rapid convergence, the $L_1$ norm (\texttt{L1Norm}) encourages sparse solutions and tends to reduce noise in the solution.
Finally, one can also use the Huber norm of the tensor field (\texttt{HuberNorm}), which acts as an \texttt{L1Norm} for large values and an \texttt{L2Norm} for small values, converging more easily than \texttt{L1Norm}.
When applied with the \gls{sh} basis-set, the \texttt{L1Norm} and \texttt{HuberNorm} are not rotational invariants, and can bias the solution towards certain directions.

In terms of optimizers \mumott{} provides gradient descent with a fixed step size (\texttt{GradientDescent}), with an option to use Nesterov accelerated momentum, as well as the LBFGS-B algorithm for quasi-Newton solution of the optimization (\texttt{LBFGS}).
For the \gls{zh} workflow (\autoref{sect:zh-basis-set}) there is both a specialized optimizer (\texttt{ZonalHarmonicsOptimizer}) and a gradient calculator (\texttt{ZonalHarmonicsGradientCalculator}).
It is a basic gradient descent optimizer with a special heuristic rule to determine a safe step size for the angle parameters.
Because of the non-convexity of the cost function, it requires a good starting guess for the angles in order to converge to a solution.

\subsection{Computational efficiency and resource requirements}
\label{sect:computational-efficiency}

The computational resources required to perform reconstructions in \mumott{} are modest compared to previous implementations due to efficient implementations of the John transform and the use of memory-efficient solvers.
The place where a user is most likely to run into problems is the memory requirement for the data set and solution vector, in addition to a few extra similarly sized arrays needed by the solvers.
The memory requirement is around a few gigabytes in the most common use cases, but increases with both the size of the voxel grid and the directional resolution.
In order to use the GPU-implementation, one requires a CUDA-compatible GPU with sufficient VRAM to store an array the size of the solution vector.

In general, the reconstruction is much less resource-hungry than the preceding data-reduction steps.
However, when conducting sweeps of regularization parameters and full $q$-resolved 3D-\gls{rsm} reconstructions, the reduced runtime from GPU-acceleration has a considerable impact.

\autoref{tab:performance_table} compares the run times for different pipelines, platforms, and configurations.
Each configuration was run \num{10} times on each platform, and the result was obtained by averaging the run times after discarding the first run, to enable on-disk caching to take place.
All runs were carried out with a maximum of \num{20} iterations, although \gls{sigtt} converged in 14 or fewer iterations in all cases.

The runs were carried out in separate, sequentially run processes, which means that just-in-time compiled kernels were not re-used beyond what is automatically cached on disk.
This has the largest effect on \gls{dd}, which creates sub-geometries for each basis function and therefore needs to re-compile code to carry out the John transform for each sub-iteration, which adds approximately a second of overhead per basis function.
Therefore, \gls{dd} can perform substantially better than what is apparent from this table when the same geometry is run multiple times in a single process, as may be done for $q$-resolved reconstruction.
The time required to load data was not included in the timing to eliminate the dependency on the file systems used for benchmarking.

\begin{table}
\caption{
    Comparison of reconstruction times in seconds averaged across 10 runs each for a typical single-$q$ dataset consisting of \num{247} projections, each with \numproduct{65x55} pixels and \num{8} detector segments, using different reconstruction pipelines and running on different computers.
    $N$ is the number of basis functions per voxel.
    In all cases, relative uncertainties were smaller than $5\%$, and are omitted to maintain ease of reading.
    The workstation (WS) data was obtained using an AMD Ryzen 7 3700X processor with 8 physical cores, \qty{64}{\giga\byte} of DDR4 \qty{2666}{\mega\hertz} RAM, and for the GPU accelerated calculations an Nvidia GeForce RTX 3060 GPU with \qty{12}{\giga\byte} of VRAM.
    The high-performance computing (HPC) CPU timings were generated using 8 top-level threads on a 64-core Intel Xeon Platinum 8358 @ \qty{2.0}{\giga\hertz} CPU with some operations utilizing lower-level multithreading.
    The HPC GPU timings were obtained used an Nvidia A100 GPU with \qty{40}{\giga\byte} of VRAM and 8 threads on 16 cores of a 64-core Intel Xeon Platinum 8358 @ \qty{2.0}{\giga\hertz}.
    }
    \label{tab:performance_table}
\begin{center}
\begin{tabular}{l*{6}r}
    \toprule
    && \multicolumn{2}{c}{CPU} && \multicolumn{2}{c}{GPU} \\
    \cmidrule{3-4} \cmidrule{6-7}
    & $N$ & WS & HPC && WS & HPC \\
    \midrule
    SIGTT
    &   6 &  23 &  18 &&   9 &  9 \\
    &  20 &  45 &  29 &&  18 & 14 \\
    &  72 & 108 &  69 &&  60 & 36 \\[6pt]
    MITRA
    &  18 &  41 &  22 &&  13 &  8 \\
    &  50 &  93 &  45 &&  40 & 14 \\
    & 162 & 271 & 156 && 115 & 37 \\[6pt]
    DD
    &  18 &  81 &  72 &&   40 &  43 \\
    &  50 & 156 & 157 &&  101 & 111 \\
    & 162 & 334 & 392 &&  290 & 346 \\[6pt]
    MOTR
    &  18 &     &     &&   9  &   8 \\
    &  50 &     &     &&  12  &  10 \\
    & 162 &     &     &&  46  &  22 \\
    \bottomrule
    \end{tabular}
\end{center}
\end{table}

\section{Outlook}
\label{sect:outlook}

Various additions and improvements to \mumott{} are foreseen for the future.
One of the main difficulties of performing tensor-tomography experiments at present in the interfacing with the existing data analysis pipelines at the various synchrotron end stations with the reconstruction pipeline. 
At present, such an integration relies on two intermediate steps.
The first is azimuthal regrouping of detector images, which results in a number of new data files containing the azimuthally regrouped intensities that are organized projection-by-projection mirroring the order in which the experiment was performed.
The second step is a slicing of the experimental dataset into the \mumott{}-compatible HDF5 files described in \autoref{tab:h5_format}, which contains the data organized by q-bins.
These extra analysis steps are often slow due to requiring many read- and write operations.
On-the-fly reconstructions would require live azimuthal regrouping of detector images and a more efficient data pipeline that allows fast slicing in the q-dimension.

At present, \mumott{} is able to compute various properties of reconstructions and to save the results to HDF5 files.
The user then has the responsibility for analysis and visualization of the reconstructed quantities.
It will be useful to add the option to write to formats compatible with common visualization software packages.

In Ref.~\citenum{nielsen_tt_2023}, simulated data was used for the purpose of validation and comparison of various reconstruction methods.
Being able to easily generate simulated data in \mumott{} would be useful not just for validation, but also to plan experiments and to generate synthetic data for training machine learning models.

The splitting of the tensor-tomography reconstruction into discrete 2D-\gls{rsm} shells is a useful simplification that reduces the size of individual reconstruction problems.
It would however often be advantageous to combine several q-bins into a single reconstruction to enforce certain types of prior knowledge of the nanostructure on the reconstruction for sample systems where an appropriate model is available.
An example of this is texture tomography \cite{frewein2024} applicable to Bragg-scattering from nanocrystalline materials where the rotational symmetries of the crystal lattice can be imposed on the reconstruction by performing a combined reconstruction of several q-bins at once.
Also in the case of fiber-scattering, different $q$-ranges can contain scattering from different sample orientations and a combined approach to reconstruction is expected to be able to alleviate missing wedge artifacts in the reconstructions. 

\begin{acknowledgments}
This work was funded by  the European Research Council Starting Grant MUMOTT (ERC-2020-StG 949301), the Swedish research council (VR 2018-041449), and the Chalmers Initiative for Advancement of Neutron and Synchrotron Techniques.
MC has received funding from the European Union’s Horizon 2020 research and innovation program under the Marie Skłodowska-Curie grant agreement No 884104.
Views and opinions expressed are those of the authors only and do not necessarily reflect those of the European Union or the European Research Council Executive Agency.
Neither the European Union nor the granting authority can be held responsible for them. 
The computations and optimizations described in this project were in part enabled by computational resources provided by Chalmers e-Commons at Chalmers University of Technology as well as the National Academic Infrastructure for Supercomputing in Sweden at NSC and C3SE partially funded by the Swedish Research Council through grant agreement No.~2022-06725.
\end{acknowledgments}

\end{document}